\documentclass[aps,prd,onecolumn,groupedaddress,showpacs,nofootinbib,amssymb]{revtex4}
\usepackage[dvips]{graphicx}
\usepackage{amssymb}
\usepackage{amsmath}
\usepackage{graphicx,color}
\usepackage{amsfonts}
\usepackage{bm}
\usepackage{cancel}
\usepackage{comment}

\newcommand\be{\begin{equation}}
\newcommand\ee{\end{equation}}

\allowdisplaybreaks[4]

\begin{document}

\title{Viable $F(R)$ Scenarios Unifying Inflation with Realistic Dynamical Dark Energy}
\author{S.D. Odintsov$^{1,2}$}
\email{odintsov@ieec.cat} \affiliation{$^{1)}$Institute of Space
Sciences (ICE, CSIC) C. Can Magrans s/n, 08193 Barcelona, Spain}
 \affiliation{$^{2)}$Instituci\'o Catalana de Recerca i Estudis Avan\c{c}ats (ICREA),
Passeig Luis Companys, 23, 08010 Barcelona, Spain}
 \author{V.K. Oikonomou$^{1,2}$}
\email{v.k.oikonomou1979@gmail.com;voikonomou@gapps.auth.gr}
\affiliation{$^{1)}$Physics Department, Observatory, Aristotle
University of Thessaloniki, Thessaloniki, Greece}
\affiliation{$^{2)}$Center for Theoretical Physics, Khazar
University, 41 Mehseti Str., Baku, AZ-1096, Azerbaijan}
\author{G.S. Sharov$^{1,2}$}
 \email{sharov.gs@tversu.ru}
 \affiliation{$^{1)}$Tver state university, Sadovyj per. 35, 170002 Tver, Russia}
 \affiliation{$^{2)}$International Laboratory for Theoretical Cosmology,
Tomsk State University of Control Systems and Radioelectronics
(TUSUR), 634050 Tomsk, Russia}

\tolerance=5000

\begin{abstract}
Two $F(R)$ gravity models are tested on the basis of their
viability during all stages of cosmological evolution. It is shown
that these models can describe both the early-time inflationary
epoch and the dark energy epoch. The models are confronted with
the latest observational data, including the Pantheon+ catalogue
with Type Ia supernovae, the Dark Energy Spectroscopic Instrument
measurements of baryon acoustic oscillations, the Hubble parameter
estimations and data from cosmic microwave background radiation.
Investigation of the viability conditions for these models, in
particular, the condition $\frac{dF}{dR}>0$ required a deep
analysis. Both models appeared to be viable during the early-time
era, but for the late-time evolution the viability conditions are
not fulfilled in definite domains in the parameter spaces of these
models. However the best fitted parameters, determined in
confrontation with the mentioned observational data, lie far from
the forbidden domains for both models. These $F(R)$ gravity models
describe the observations with the large advantage over the
$\Lambda$-Cold-Dark-Matter model, not only in $\chi^2$ statistics,
but also with Akaike and Bayesian information criteria. This
success of the two $F(R)$ gravity scenarios is connected with
their capability to mimic dynamical dark energy, similarly to
models with variable equation of state, that is necessary for
describing the latest Pantheon+ and DESI observational data.
\end{abstract}


\maketitle

\section{Introduction}

Progress in cosmology always was supported by incoming
observational data, there were some periods in recent scientific
history, when new observations led to radical transformations in
the cosmological landscape. One of the most bright examples of
such revolutionary changes, followed after measurements of Type Ia
supernovae (SNe Ia) parameters in 1998\,--\,1999
\cite{Riess:1998,Perlmutter:l999}. These datasets and further
observations bore witness to the accelerated expansion of the
Universe at the late-time epoch driven by a negative pressure
fluid dubbed dark energy. This picture was described in numerous
cosmological scenarios with a leading dark energy fraction
nowadays, and the most successful among them invariably was the
$\Lambda$-Cold-Dark-Matter model ($\Lambda$CDM) with the
cosmological constant $\Lambda$ generating the dark energy
evolution, for reviews on the subject, see Refs.
\cite{Peebles:2003,BambaCNO:2012,reviews2,reviews3,reviews4,reviews5}.

The $\Lambda$CDM model encounters some theoretical and
observational problems: the physical nature of its main components
is unknown, that is dark energy and cold dark matter are still a
mystery. Furthermore, other problems of the $\Lambda$CDM model
are, the coincidence problems with close fractions of these
components nowadays, the fine-tuning problem for $\Lambda$
\cite{Peebles:2003,BambaCNO:2012}, the Hubble constant tension
between $\Lambda$CDM-based early-Universe estimations of $H_0$
from Cosmic Microwave Background radiation (CMB)
\cite{Planck2018}, and local distance-ladder measurements by the
SH0ES collaboration \cite{Riess2021,DiValentino:2020naf}. Many
authors tried to solve these problems and tensions in alternative
cosmological scenarios, including interacting dark components and
other modifications of General Relativity
\cite{DiValentino:2020naf,Dai:2020rfo,He:2020zns,Nakai:2020oit,Agrawal:2019dlm,Yang:2018euj,
Ye:2020btb,Vagnozzi:2021tjv,Desmond:2019ygn,Hogas:2023pjz,OColgain:2018czj,Vagnozzi:2019ezj,
Krishnan:2020obg,Colgain:2019joh,Vagnozzi:2021gjh,Lee:2022cyh,Krishnan:2021dyb,Ye:2021iwa,Ye:2022afu,Verde:2019ivm,Menci:2024rbq,Adil:2023ara,Reeves:2022aoi,Ferlito:2022mok,
Vagnozzi:2021quy,DiValentino:2020evt,DiValentino:2019ffd,DiValentino:2025sru,Valletta:2025bgu,Montani:2025nmz,Fazzari:2025mww,Montani:2025jkk,Schiavone:2024heb,Montani:2024pou,Montani:2024ntj,Escamilla:2024xmz,Montani:2023ywn,Montani:2023xpd}
with achievements in some directions, but the $\Lambda$CDM model
kept its leading position in statistically analyzed description of
all the available observational data.

However, during the last two years this landscape suffers from
serious groundbreaking evidence challenging the validity of the
$\Lambda$CDM model. The latest observational data, in particular,
the SNe Ia datasets from the Pantheon+ and Union3 catalogues
\cite{PantheonP:2022,Union3:2023} and the Dark Energy
Spectroscopic Instrument (DESI) measurements
\cite{DESI:2024,DESI:2025zgx} of Baryon Acoustic Oscillations
(BAO) led to some essential transformations in cosmology, in
particular, the dominating position of the $\Lambda$CDM model was
questioned. The mentioned observational data can be described more
successfully in numerous models with a dynamical dark energy or
variable equation of state (EoS) for dark energy
\cite{Cai:2025mas,Ye:2024ywg,Chaudhary:2025vzy,Chaudhary:2025uzr,OdintsovSGS_game:2024,Giare:2024smz,Pan:2025qwy,Yang:2025mws,Zhang:2025bmk,Ong:2025utx,Nojiri:2025uew}.
Analysis of DESI DR2 \cite{DESI:2025zgx} and other observations
indicated that the dark energy EoS evolves from a phantom to a
quintessence EoS during the late-time epoch. Note that the models
with dynamical dark energy confronted with the newest
observational data have advantages in comparison with the
$\Lambda$CDM scenario, if information criteria are used in
statistical analysis.

Dark energy in different forms, can behave in a dynamically
evolving way, not only in models with a variable EoS, but also in
modified gravity theories
\cite{BambaCNO:2012,reviews2,reviews3,reviews4,reviews5}. In
particular, this behavior of dynamical dark energy occurs in
$F(R)$ gravity theories, which contain non-trivial dependence on
the Ricci scalar $R$ in the gravitational Lagrangian
\cite{Nojiri:2003ft,Capozziello:2005ku,Hwang:2001pu,Song:2006ej,Faulkner:2006ub,Olmo:2006eh,
Sawicki:2007tf,Faraoni:2007yn,Carloni:2007yv,Nojiri:2007as,Deruelle:2007pt,Appleby:2008tv,
Dunsby:2010wg,Hu:2007nk,Bamba:2012qi,inflation5,Linder:2009,OnofrioOS:2025,
OdintsovSGS:2017,OdintsovSGSlog:2019,OdintsovSGStens:2021,OdintsovSGS_Axi:2023,
OdintsovSGS_LnAx:2024,CognolaENOSZ:2008,ElizaldeNOSZ:2011,
Oikonomou:2025qub,OdintsovOS:2025,Odintsov:2025kyw,CaiS:2014,NunesPSA:2016,Chen:2019}.
These models can successfully unify the early-time inflationary
era and the late-time dark energy epoch.

In our analysis we include two $F(R)$ scenarios, chosen from the
more wide class of viable $F(R)$ gravity models, suggested earlier
in Refs.~\cite{Oikonomou:2025qub,OdintsovOS:2025}. These
scenarios, named as ``logarithmic model'' and ``model with an
exponent'', have the $\Lambda$CDM-like asymptotic behavior at
early times (or at large $R$), and we demonstrate that they mimic
dynamical dark energy at late times. These models prove to be more
successful than $\Lambda$CDM in describing the observational data,
including the BAO DESI data 2025. We also test the viability
conditions, in particular, the condition $\frac{dF}{dR}>0$ for
these $F(R)$ scenarios during all cosmological history.

The article is organized as follows: in section \ref{Dynamics},
the dynamical equations for $F(R)$ gravity are described and
adopted for further late-time analysis, including the viability
conditions. In section \ref{Log} the logarithmic $F(R)$ model is
investigated with its viability and observational tests with SNe
Ia, $H(z)$, CMB and BAO DESI data. The same analysis and
statistical calculations are performed for the exponential $F(R)$
gravity model. Finally, the conclusions along a discussion on the
results follow at the end of the article.

\section{ $F(R)$ Gravity Framework, Dynamical Evolution and Viability}
\label{Dynamics}

The $F(R)$ gravity theory in the presence of perfect matter fluids
has the following action,
\begin{equation}
\label{action1}
\mathcal{S}=\int{d^4x\sqrt{-g}\left(\frac{F(R)}{2\kappa^2}+\mathcal{L}_m\right)}\, ,
\end{equation}
where $\kappa^2=8\pi G$, with $G$ being the Newtonian
gravitational constant, $\mathcal{L}_m$ is the Lagrangian density
of the perfect matter fluid components. $F(R)$ gravity models can
unify the inflationary era with the dark energy epoch within the
same theoretical framework. This unification may be achieved, in
particular, if $F(R)$ contains the terms
\cite{CognolaENOSZ:2008,ElizaldeNOSZ:2011,Oikonomou:2025qub,OdintsovOS:2025}
\begin{equation} \label{FR3terms}
      F(R)=R+f(R)=R+F_\mathrm{inf}(R)+F_\mathrm{DE}(R)\,,
\end{equation}
 where $F_\mathrm{inf}(R)$ and $F_\mathrm{DE}(R)$ are the inflationary and the dark energy
terms respectively. The inflationary term is used below in the
form $F_\mathrm{inf}={R^2}/{M^2}$, where the constant $M\sim
3\cdot10^{22}$ eV is assumed to be large enough to make the term
$F_\mathrm{inf}$ negligible near and after the recombination
epoch, at redshifts $0\le z\le 10^4$,

We fix the background spacetime used below in this article to be
the spatially flat Friedmann-Robertson-Walker (FRW) metric,
\begin{equation}
 ds^2 = - dt^2 + a(t)^2 \sum_{i=1,2,3} \left(dx^i\right)^2\, ,
\label{FRW}\end{equation}
 where $a(t)$ is the scale factor, $H=\dot{a}/a$ is the Hubble parameter, the ``dot'' indicates differentiation with respect to the cosmic time.

The field equations deduced from the  action (\ref{action1}) may
be rewritten in the Einstein-Hilbert form \cite{OdintsovOS:2025}:
\begin{eqnarray}
 3 H^2&=& \kappa^2 \rho_\mathrm{tot}\,,
\label{Friedtot} \\
-2 \dot H&=&\kappa^2 (\rho_\mathrm{tot} + P_\mathrm{tot})\,.
   \label{Raychtot}
\end{eqnarray}
Here the total energy density and total pressure are,
\begin{equation}\label{rhoP}
\rho_\mathrm{tot}=\rho_m +\rho_r+\rho_{DE}\,,\qquad
 P_\mathrm{tot}=P_m + P_r +P_{DE}\,
\end{equation}
and include contributions from  the cold matter ($\rho_m$,
$P_m=0$), from radiation ($\rho_r$, $P_r$) and the geometric part,
generated by $F(R)$ gravity:
\begin{eqnarray}\label{rDE}
  \kappa^2  \rho_{DE}&=&\frac{F_R R - F}{2} + 3 H^2 (1-F_R)-3H \dot F_R \, ,\\
\label{PDE}
    P_{DE}&=&\frac{\ddot F -H \dot F +2 \dot H (F_R -1)}{\kappa^2} - \rho_{DE} \,
    ,
\end{eqnarray}
where $F_R = \frac{\partial F}{\partial R}$.

In this paper, we investigate two $F(R)$ gravity scenarios,
studied previously in
Refs.~\cite{Oikonomou:2025qub,OdintsovOS:2025}. The first model
has the $\log R$ factor in its $F_\mathrm{DE}$ term:
\begin{equation}\label{fr221}
    F(R)=R+\frac{R^2}{M^2}-\frac{\beta  \Lambda }{\gamma +{1}\big/{\log \left(\epsilon \frac{R}{m_s^2}\right)}},
\end{equation}
with $\beta$, $\Lambda$, $\gamma$ and $\epsilon$ being positive
constants, $m_s^2=\frac13 \kappa^2 \rho_m^0$. We focus on $F(R)$
gravity scenarios with a $\Lambda$CDM-like asymptotic behavior in
the large $R$ limit. More precisely, the considered scenarios have
limiting behavior similar to the $\Lambda$CDM Lagrangian,
\begin{equation}\label{LCDM}
F(R)=R -2\Lambda
 \end{equation}
at the epoch, when the Ricci scalar $R$ is much larger compared to
the cosmological constant $\Lambda$, bur $R$ is much less than its
value $R_i$ at the beginning of the  inflationary era, so we can
neglect the term $F_\mathrm{inf}(R)$.

One can see that the model (\ref{fr221}) tends to the $\Lambda$CDM
Lagrangian (\ref{LCDM}) at $R\gg m_s^2/\epsilon$,  if
$\beta/\gamma=2$ and  $\Lambda$ is the same cosmological constant.
If we assume $\beta=2\gamma$ and denote
$\alpha=\epsilon\cdot2\Lambda/m_s^2$, the Lagrangian (\ref{fr221})
can be rewritten as follows,
\begin{equation}\label{ModLog}
    F(R)=R+\frac{R^2}{M^2}-2\Lambda\bigg[1-\frac1{1+\gamma\log
    \left(\alpha\frac{R}{2\Lambda}\right)}\bigg]\,.
\end{equation}
The last term $F_\mathrm{DE}(R)$ tends to the $\Lambda$CDM limit
$-2\Lambda$ if $R\to\infty$, but also at $\gamma\to\infty$ and
$\alpha\to\infty$. Below the scenario (\ref{ModLog}) will be named
as the ``logarithmic'' model.

The second scenario under consideration has the following $F(R)$
function \cite{Oikonomou:2025qub,OdintsovOS:2025},
\begin{equation}\label{simexp}
F(R)=\mu R+\frac{R^2}{M^2}+\lambda
R\,e^{\epsilon\left(\frac{\Lambda}{R}\right)^{\beta}}+\nu \Lambda \, ,
\end{equation}
with $\epsilon$, $\mu$, $\lambda$, $\beta$ and $\nu$ being dimensionless parameters. The
non-inflationary part $F(R)=R+F_\mathrm{inf}(R)$ will tend to the $\Lambda$CDM limit
$R-2\Lambda$ at $R\to\infty$ if $\mu=1-\lambda$ and the suitable choice of $n$. The
parameter $\mu=1-\lambda$ is a measure of ``mixing'' between $\Lambda$CDM and this
$F(R)$ scenario: in particular, in the case $\mu=1$, $\lambda=0$ the  Lagrangian
(\ref{simexp}) takes the pure $\Lambda$CDM form (\ref{LCDM}). So the most interesting is
the opposite case $\mu=0$, $\lambda=1$, considered further, with
\begin{equation}\label{ModExp}
F(R)=\frac{R^2}{M^2}+
R\,\exp\bigg[\,\varepsilon\Big(\frac{2\Lambda}{R}\Big)^{\beta\,}\bigg]+\nu
\Lambda \, ,\qquad \nu=\left\{\begin{array}{ll}-2,& \beta>1,\\
-2-2\varepsilon, & \beta=1.\end{array} \right.
\end{equation}
Here $\varepsilon=\epsilon\cdot2^{-\beta}$, the constant $\nu$  is
responsible for the $\Lambda$CDM limit (\ref{LCDM}) at
$R\gg\Lambda$, this limit exists if $\beta\ge 1$.

Note that both the scenarios we consider in this article, have two
additional dimensionless free parameters related with their
late-time dynamics: they are $\gamma$ and $\alpha$ for the
logarithmic model (\ref{ModLog}) and $\varepsilon$ and $\beta$ for
the exponential model (\ref{ModExp}). The early-time inflationary
dynamics is also controlled by the parameter $M$.

During all stages of evolution any $F(R)$ scenario should satisfy the viability
conditions
\cite{Hu:2007nk,CognolaENOSZ:2008,ElizaldeNOSZ:2011,Chen:2019,Oikonomou:2025qub,OdintsovOS:2025}
including  the inequalities
\begin{equation}\label{FRFRR}
F_R>0 \, ,\qquad F_{RR} > 0\,,
\end{equation}
where the first condition is necessary to avoid anti-gravity
effects and the inequality $F_{RR} > 0$ supplies stability of the
cosmological perturbations during the matter dominated era and
compatibility with local solar system tests. We should add the
viability conditions related with inflationary and
post-inflationary dynamics of a $F(R)$ gravity, more precisely,
the requirements of a stable de Sitter point existence and
non-negativity of the scalaron mass
$$
m^2=\frac{1}{3}\left(-R+\frac{F_R}{F_{RR}} \right)\, ,
$$
measuring de Sitter perturbations. If we add the requirement for $m^2$ to grow with
growing $R$, these conditions are reduced to \cite{Oikonomou:2025qub,OdintsovOS:2025}
\begin{equation}\label{yxineq}
0< y \leq 1\,,\qquad  x\leq 0\,,
\end{equation}
where
\begin{equation}\label{yx}
 y=\frac{R\,F_{RR}}{F_R}\,,\qquad  x= 4\frac{R F_{RRR}}{F_{RR}}\,.
\end{equation}
The condition $y>0$ is the consequence of the inequalities (\ref{FRFRR}), it should be
fulfilled at all times. This is also true for the restriction $x\leq 0$. But the
condition $y \leq 1$ is related only with existence of a stable de Sitter solution, in
particular, during the inflationary era.

Table \ref{T1} presents the mentioned functions $F_R$, $F_{RR}$, $x(R)$ for the
logarithmic $F(R)$ model (\ref{ModLog}) and for the model (\ref{ModExp}) with the
exponent to analyze their viability. For brevity we use the normalized dimensionless
Ricci scalar
\begin{equation}\label{calR}
{\cal R}=\frac{R}{2\Lambda}
\end{equation}
 and the following notation:
\begin{equation}\label{ellR}
\ell_R=1+\gamma\log(\alpha{\cal R})\,.
\end{equation}
\begin{table}[hb]
  \begin{center}
    \caption{Parameters for the models (\ref{ModLog})   and (\ref{ModExp}) to test their viability.}
    \label{T1}
    \begin{tabular}{|c|c|c|}
    \hline
      {Parameter} & Log $F(R)$ model (\ref{ModLog}) & Model (\ref{ModExp}) with $e^{\varepsilon{\cal R}^{-\beta}}$
      \rule{0pt}{1.2em} \\  \hline
      $F_R$  & $1+2RM^{-2}-\gamma/({\cal R}\ell_R^2)$  &
       $2RM^{-2}+(1-\varepsilon\beta{\cal R}^{-\beta})\,e^{\varepsilon{\cal R}^{-\beta}}$\rule{0pt}{1.3em} \\  
      $F_{RR}$ &$2M^{-2}+2\Lambda\gamma\dfrac{2\gamma+\ell_R}{R^2\ell_R^3}$&
       $\dfrac2{M^2}+\dfrac{\varepsilon\beta}R\big[(\beta-1){\cal R}^{-\beta}+\varepsilon\beta{\cal R}^{-2\beta}\big]\,
       e^{\varepsilon{\cal R}^{-\beta}}$ \rule{0pt}{1.7em} \\  
      $x(R)$  &  $-\dfrac{8\gamma[\ell_R+3\gamma(1+\gamma/\ell_R)]}{\gamma(2\gamma+\ell_R)+R^2\ell_R^3/(\Lambda M^2)}$
        & $\;-4\dfrac{(\beta^2-1){\cal R}^{-\beta}+3\varepsilon\beta^2{\cal R}^{-2\beta}+\varepsilon^2\beta^2{\cal R}^{-3\beta}}
        {(\beta-1){\cal R}^{-\beta}+\varepsilon\beta{\cal R}^{-2\beta}+2e^{-\varepsilon{\cal R}^{-\beta}}R/(M^2\varepsilon\beta)} $  \rule[-0.1em]{0pt}{1.9em} \\  \hline
    \end{tabular}
  \end{center}
\end{table}
Recall that the model parameters  $\gamma$, $\alpha$, $\epsilon$
are positive and $\beta$ is limited as $\beta\ge 1$. In this case
one may conclude from Table~\ref{T1} that the viability conditions
$F_{RR}>0$ and $x<0$ are fulfilled for the model (\ref{ModExp})
with the exponent during all stages of evolution. However, the
conditions $F_{R}>0$ and $y>0$ for this model may be violated at
late times (when ${\cal R}$ is not large) if the term
$\varepsilon\beta{\cal R}^{-\beta}$ appears to be larger than
unity.

For the logarithmic $F(R)$ model (\ref{ModLog}) the conditions
(\ref{FRFRR}) and (\ref{yxineq}) need an additional verification,
because they can be violated if the factor (\ref{ellR})
$\ell_R=1+\gamma\log(\alpha{\cal R})$ becomes too small or
negative. This potential violation is also related with late
times, small values of $\alpha$ and large $\gamma$. In particular,
in $F_R$ the negative term  $-\gamma/({\cal R}\ell_R^2)$ can
dominate at small values of ${\cal R}$, that leads to the
forbidden inequality $F_R<0$.

However, to verify  the conditions (\ref{FRFRR}) at late times, we should know how the
Ricci scalar $R$ evolves in a considered scenario, more precisely, know the lowest value
$R_\mathrm{min}=\min R$ of $R$ during its evolution. To determine this evolution, we
should fix not only the mentioned parameters $\gamma$, $\alpha$ (or $\epsilon$ and
$\beta$ for the second model), but also other model parameters, in particular, the
fraction of cold matter density and the $\Lambda$ term fraction:
 \begin{equation}
\Omega_m^0=\frac{\kappa^2 \rho_m^0}{3H_0^2}\,,\qquad
\Omega_\Lambda=\frac{\Lambda}{3H_0^2}\,.
 \label{Omega_mL}\end{equation}
Here, as usual,  $H_0=H(t_0)$ is the Hubble constant,
$\rho_m^0=\rho_m(t_0)$ is the cold matter energy density nowadays
(at $t=t_0$). Due to this reason our analysis of the viability
conditions (\ref{FRFRR}), (\ref{yxineq}) will follow after some
details of $F(R)$ dynamics in the next section (see Fig.~\ref{F1}
below). Note here that at early times at the limit ${\cal
R}\to\infty$, the negative terms $-\gamma/({\cal R}\ell_R^2)$ and
$-\varepsilon\beta{\cal R}^{-\beta}$ in $F_R$ tend to zero for
both models, so all conditions (\ref{FRFRR}) and (\ref{yxineq})
are satisfied.

\smallskip

The equations (\ref{Friedtot}), (\ref{Raychtot}) of $F(R)$ gravity
models can be reduced to the system of equations
\cite{OdintsovSGS:2017,OdintsovSGSlog:2019,OdintsovSGStens:2021,OdintsovSGS_Axi:2023,OdintsovOS:2025}:
 \begin{eqnarray}
\frac{dH}{d\log a}&=&\frac{R}{6H}-2H\ , \label{eqH1} \\
\frac{dR}{d\log
a}&=&\frac1{F_{RR}}\bigg(\frac{\kappa^2\rho}{3H^2}-F_R+\frac{RF_R-F}{6H^2}\bigg)\ .
 \label{eqR1}
 \end{eqnarray}
The first equation is equivalent to the relation $R=6\dot H +
12H^2$. In this paper, we integrate numerically the system
(\ref{eqH1}), (\ref{eqR1}) for a chosen $F(R)$ model with
$\Lambda$CDM-like  behavior at high $R$ using the approach
developed previously in papers
\cite{OdintsovSGS:2017,OdintsovSGSlog:2019,OdintsovSGStens:2021,OdintsovSGS_Axi:2023,OdintsovOS:2025,Odintsov:2025kyw}.
In this approach we integrate the equations with growing $a$ (to
the future direction) starting from some initial point
$a_\mathrm{ini}$ with initial conditions assuming a
$\Lambda$CDM-like asymptotic behavior at and before
$a_\mathrm{ini}$. This initial point is determined from the
condition of the defined small value for the term
$F_{RR}(a_\mathrm{ini})$ in the right hand side of
Eq.~(\ref{eqR1}). Recall that at high values of the curvature $R$,
the dimensionless expression $2\Lambda F_{RR}$ for the considered
models (\ref{ModLog}) and (\ref{ModExp}) tends to
$4\Lambda/M^2\sim$, as can be seen in Table~\ref{T1}. This value
is extremely small: $4\Lambda/M^2\sim10^{-110}$
\cite{OdintsovOS:2025}. Hence, in our calculations we should
assume that at high $R$ the denominator $F_{RR}$  in the right
hand side of Eq.~(\ref{eqR1}) tends to zero, so viable solutions
will exist if the corresponding numerator tends to zero too.

These viable solutions should have a $\Lambda$CDM-like asymptotic
behavior at $R\to\infty$ or at $a<a_\mathrm{ini}$ with the Hubble
parameter  $H(a)$ and the Ricci scalar $R(a)$  in the form
\cite{OdintsovSGS:2017,OdintsovSGSlog:2019,OdintsovSGStens:2021,OdintsovSGS_Axi:2023,OdintsovSGS_game:2024,OdintsovOS:2025}:
 \begin{equation} \label{asymLCDM}
 \frac{H^2}{H^{*2}_0}=\Omega_m^{*} \big(a^{-3}+ X_r a^{-4}\big)+\Omega_\Lambda^{*}\,,\qquad
 \frac{R}{2\Lambda}=2+\frac{\Omega_m^{*}}{2\Omega_\Lambda^{*}}a^{-3}\ .
 \end{equation}
Here we introduce the $\Lambda$CDM-asymptotical Hubble constant
$H^*_0$  at the initial point $a_\mathrm{ini}$ that differs from
the true Hubble constant $H_0=H(t_0)$ achieved during evolution of
a chosen $F(R)$ model from $a_\mathrm{ini}$ to the present day
value $a=1$. The value $H^*_0$ determines the parameters,
\begin{equation} \label{Omega_mL2}
\Omega_m^{*}=\frac{\kappa^2\rho_m^0}{(H_0^*)^2}\,,\qquad
\Omega_\Lambda^*=\frac{\Lambda}{3(H_0^*)^2}\,,
 \end{equation}
they are analogs of the standard $\Omega_m^0$, $\Omega_\Lambda$
(\ref{Omega_mL}) and connected with them as follows,
 \begin{equation} \label{H0Omm}
 \Omega_m^0H_0^2=\Omega_m^{*}(H^{*}_0)^2=m_s^2\ ,
 \qquad  \Omega_\Lambda H_0^2=\Omega_\Lambda^{*}(H^{*}_0)^2=\frac{\Lambda}3\ .
 \end{equation}
The present day radiation to matter ratio in Eq.~(\ref{asymLCDM})
is,
   \begin{equation} \label{Xrm}
X_r=\frac{\rho_r^{0}}{\rho_m^{0}}=2.9656\cdot10^{-4}
\end{equation}
which is fixed from Planck data
\cite{OdintsovSGStens:2021,OdintsovSGS_Axi:2023,OdintsovOS:2025}.
Further details of the dynamics of the $F(R)$ gravity models under
study, are considered in the next sections.

\section{Logarithmic $F(R)$ Gravity Model}
\label{Log}

The initial point of integration $a_\mathrm{ini}$ is determined
from the following condition
\cite{OdintsovSGStens:2021,OdintsovSGS_Axi:2023}: the
dimensionless term $2\Lambda F_{RR}$ in the denominator of the
right hand side of Eq.~(\ref{eqR1}) (it tends to zero at
$R\to\infty$) should be equal to a small value $\delta$ of order
$10^{-10}$, and $R$ has the $\Lambda$CDM-like asymptotic form
(\ref{asymLCDM}). For the logarithmic model (\ref{ModLog})
$a_\mathrm{ini}$ may be calculated from two equations
 \begin{equation} \label{ainiL}
2\Lambda F_{RR}=\gamma\frac{2\gamma+\ell_{R_\mathrm{ini}}}{{\cal
R}_\mathrm{ini}^2\ell_{R_\mathrm{ini}}^3}=\delta\;,
 \qquad a_\mathrm{ini}= \bigg[\frac{2\Omega_\Lambda^*}{\Omega_m^*}({\cal R}_\mathrm{ini}-2)\bigg]^{-1/3}\;,
 \end{equation}
Here ${\cal R}_\mathrm{ini}$ is determined from the first equation
with $\ell_{R_\mathrm{ini}}=1+\gamma\log(\alpha{\cal
R}_\mathrm{ini})$. Starting from this $a_\mathrm{ini}$ we
integrate the system of  equations (\ref{eqH1}), (\ref{eqR1}) that
can be rewritten for the logarithmic $F(R)$ model (\ref{ModLog})
in the form,
\begin{eqnarray}
\frac{dE}{d\log a}&=&\Omega_\Lambda^{*}\frac{{\cal R}}{E}-2E\,,\qquad\quad E=\frac{H}{H_0^{*}},\label{eqH2} \\   
\frac{d{\cal R}}{d\log a}&=&\frac{{\cal R}^2\ell_R}{\gamma(2\gamma+\ell_R)}
\bigg[\frac{\Omega_m^{*}(a^{-3}+ X_r a^{-4})\ell_R^2
+\Omega_\Lambda^{*}\gamma\big(\ell_R\log(\alpha{\cal R})-1\big)}{E^2}
-\ell_R^2+\frac\gamma{\cal R}\bigg]\,,
  \label{eqR2}
  \end{eqnarray}
where we used as dimensionless variables the normalized  Hubble
parameter $E=H/{H_0^{*}}$ and the Ricci scalar ${\cal R}$
(\ref{calR}).

Integrating numerically this system of equations with the initial conditions
(\ref{asymLCDM}) at  $a_\mathrm{ini}$ we obtain the solution $E=E(a)$, ${\cal R}={\cal
R}(a)$ for any set of model parameters $\gamma$, $\alpha$, $\Omega^*_m$,
$\Omega^*_\Lambda$ or $\gamma$, $\alpha$, $\Omega^0_m$, $\Omega_\Lambda$, because the
last two parameters can be recalculated via Eqs.~(\ref{H0Omm}) and the relation
$E|_{a=1}=H_0/H_0^{*}$, coming from the definition of $E$:
 \begin{equation} \label{Omm2}
 \Omega_m^0=\Omega_m^{*}/(E|_{a=1})^2\,,
 \qquad  \Omega_\Lambda =\Omega_\Lambda^{*}/(E|_{a=1})^2\,.
 \end{equation}
This approach gives possibilities to solve the mentioned above
viability problem with the conditions (\ref{FRFRR}),
(\ref{yxineq}). For this purpose we should know limits of the
Ricci scalar evolution  ${\cal R}(a)$. However, for the model
(\ref{ModLog}) this evolution not only depends on the parameters
$\gamma$, $\alpha$ from the Lagrangian (\ref{ModLog}), but also on
the intrinsic parameters $\Omega^0_m$ and $\Omega_\Lambda$. Due to
this reason we should solve the viability problem simultaneously
with testing this model in confrontation with observational data.
These tests will give the best fitted values of all model
parameters and domains of their suitable values.

In this paper, we test the considered $F(R)$ models (\ref{ModLog})
and (\ref{ModExp}) with the following observational data:  the
Pantheon+ catalog of Type Ia supernovae (SNe Ia)
\cite{PantheonP:2022} and also baryon acoustic oscillations (BAO)
data from the DESI 2025 \cite{DESI:2025zgx}, the Hubble parameter
measurements $H(z)$ or Cosmic Chronometers (CC) and the Planck
data from CMB measurements \cite{Planck2018}.

For this purpose we solve the system (\ref{eqH2}), (\ref{eqR2})
with a set of model parameters, including the Hubble constant
$H_0$, we obtain the Hubble parameter $H(a)$ or $H(z)$, expressed
via the redshift,
\begin{equation}
  z=\frac{1}{a}-1\, ,
\end{equation}
and we calculate the $\chi^2$ functions $\chi^2_\mathrm{SN}$,
$\chi^2_\mathrm{BAO}$, $\chi^2_H$,  $\chi^2_\mathrm{CMB}$
\cite{OdintsovSGS_game:2024,OdintsovSGS_Axi:2023,OdintsovOS:2025,Odintsov:2025kyw}.
These $\chi^2$ functions and corresponding data are described in
the Appendix.

\begin{figure}[th]
   \centerline{ \includegraphics[scale=0.68,trim=5mm 0mm 2mm -1mm]{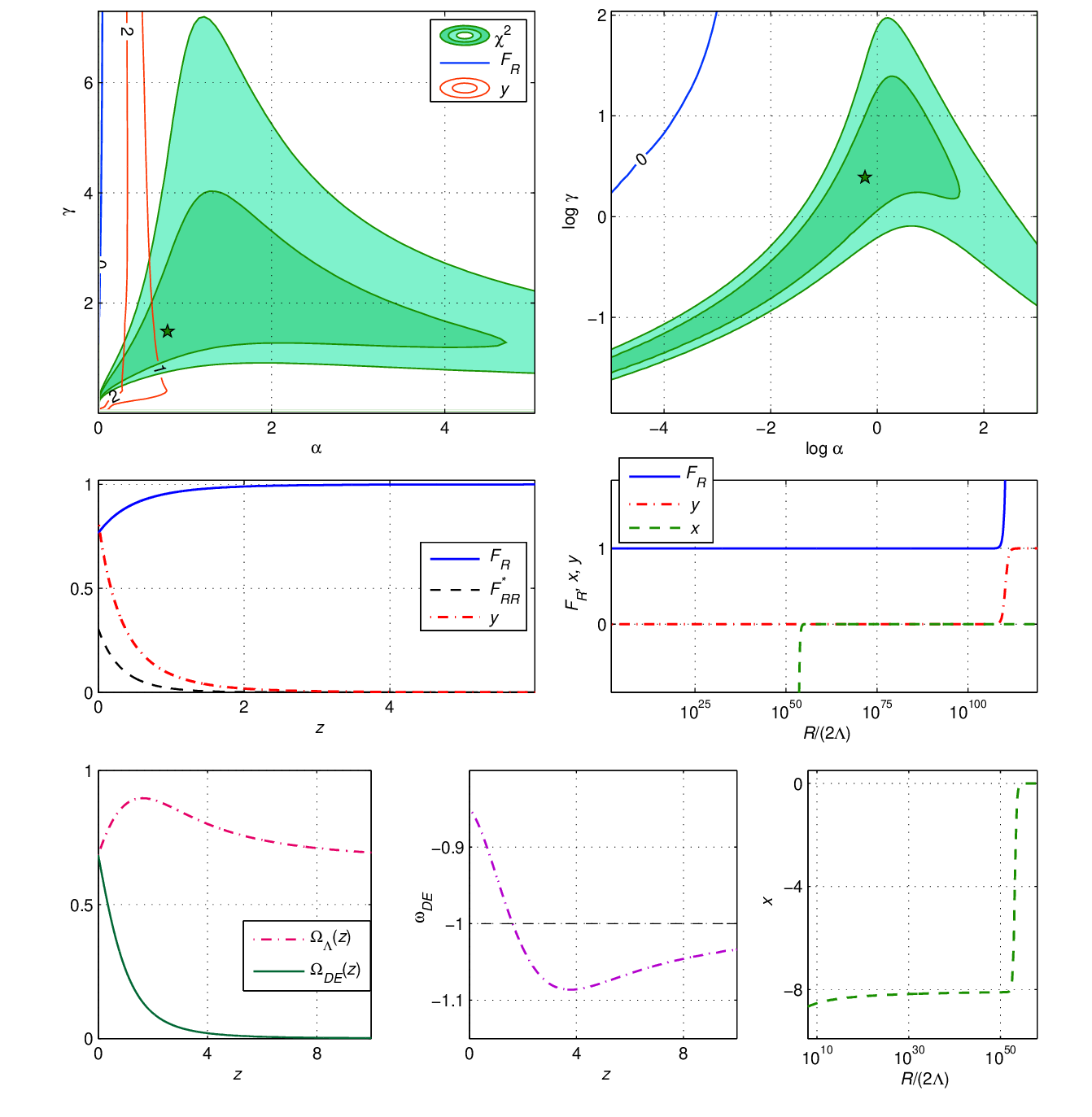}}
\caption{Contour plots of $\chi^2$ with $1\sigma$, $2\sigma$ CL in the $\alpha-\gamma$
plane for the logarithmic model (\ref{ModLog}) (the top panels); evolution of $F_R$,
$F_{RR}^*=2\Lambda F_{RR}$, $y$ and $x$ az functions of redshift $z$ (late-time) and the
Ricci scalar ${\cal R}=R/(2\Lambda)$ (early-time dynamics) in the middle panels; the
dark energy density and EoS parameters in the bottom panels. }
  \label{F1}
\end{figure}

We seek the best fit parameters of a considered $F(R)$ scenario
minimizing the total $\chi^2$ function,
 \begin{equation}
  \chi^2=\chi^2_\mathrm{SN}+\chi^2_H+\chi^2_\mathrm{CMB}+\chi^2_\mathrm{BAO}\ .
 \label{chitot} \end{equation}
The results of this $\chi^2$ function calculation in the
$\alpha-\gamma$ plane for the logarithmic model (\ref{ModLog}) and
also testing its viability with the conditions (\ref{FRFRR}),
(\ref{yxineq}) are presented in Fig.~\ref{F1}. The contour plots
in the top panels correspond to $1\sigma$ (68.27\%) and $2\sigma$
(95.45\%) confidence regions for the two-parameter distribution
$$\chi^2(\alpha,\gamma)=\min\limits_{\Omega_m^0,\Omega_\Lambda,H_0}\chi^2(\alpha,\gamma,\Omega_m^0,\Omega_\Lambda,H_0)\;.$$
The stars denote the best fits where $\chi^2$ achieves its
minimum. The best fits with $1\sigma$ errors for all free model
parameters may also be seen in Fig.~\ref{F2} and in
Table~\ref{TMod} below.
The contour plots in the $\alpha-\gamma$ plane are drawn in the
top-left panel of Fig.~\ref{F1}  and in the top-right panel we use
a more convenient logarithmic scale for the same contours. The
blue lines in the top panels correspond to the equality
$\min\limits_R F_R=0$, where $F_R$ is calculated at a certain
point $(\alpha,\gamma)$ with the best fitted $\Omega_m^0$ and
$\Omega_\Lambda$. Thus, the blue lines are borders of the domain
with small $\alpha$ and large $\gamma$, where $\min F_R<0$ and the
(anti-gravity) viability condition (\ref{FRFRR})  $F_R>0$  is
violated. Recall that at small $\alpha$ and large $\gamma$ the
factor (\ref{ellR}) $\ell_R=1+\gamma\log(\alpha{\cal R})$ can be
close to zero and lead to the mentioned violation. However, one
can see in the top panels of Fig.~\ref{F1} that the domain with
the violation $\min F_R<0$ is situated far from the best fit
values of $\alpha$ and $\gamma$.

Note that the second  viability condition (\ref{FRFRR}) $F_{RR}>0$
is fulfilled in all the $\alpha-\gamma$ plane. But in the
mentioned domain with $\min F_R<0$ the condition (\ref{yxineq})
$y>0$ for the parameter (\ref{yx}) $y=R\,F_{RR}/{F_R}$ is violated
too. The second condition (\ref{yxineq}) $y\le1$ is violated at
late times even for close to the best fits values of $\alpha$ and
$\gamma$, as can be seen in the top-left panel of Fig.~\ref{F1},
where the lines $\max y=1$ and $\max y=2$ are drawn. However, the
condition $y\le1$ should be fulfilled only at and near de Sitter
stage of expansion, in particular, during the inflationary era. In
the middle-right panels of Fig.~\ref{F1} we see that for the best
fit solution the condition $y\le1$ is fulfilled at early times,
where $z$ and $R$ are large.

The late and early time behavior of the parameters $F_R$,
$F_{RR}^*=2\Lambda F_{RR}$, $y$ and $x$ for the best fit solution
of the model (\ref{ModLog}) (with model parameters from
Table~\ref{TMod}) is shown in the middle panels of Fig.~\ref{F1}.
The middle-left panel illustrates the late-time dynamics of these
parameters as functions of redshift $z$. The viability conditions
(\ref{FRFRR}) and (\ref{yxineq}) are fulfilled in this redshift
range and earlier for the considered best fit solution. The plot
for $x(z)$ is not shown, this value is less than $-8$ during all
late-time evolution (see the bottom-right panel), that satisfies
the condition (\ref{yxineq}) $x<0$.

The middle-right and bottom-right panels of  Fig.~\ref{F1} present
the early-time dynamics of the mentioned parameters $F_R$, $y$ and
$x$ as functions of the normalized Ricci scalar ${\cal
R}=R/(2\Lambda)$. The plot $F_{RR}^*({\cal R})=2\Lambda F_{RR}$ is
not shown because this expression is positive and very small at
high ${\cal R}$. In particular, it lies in the range
$0<F_{RR}^*({\cal R})<10^{-20}$, if ${\cal R}>10^9$. In the early
times, the inflationary term $F_\mathrm{inf}={R^2}/{M^2}$ in
$F(R)$ or the corresponding term
$2R/{M^2}=\frac{4\Lambda}{M^2}{\cal R}$ in
 $$
 F_R= \frac{4\Lambda}{M^2}{\cal R}+ 1 -\frac\gamma{{\cal R}\ell_R^2}
 $$
plays its important role if ${\cal R}$ is of order or larger than the value
\begin{equation}
   {\cal R}_\mathrm{inf}=\frac{M^2}{4\Lambda}\approx1.89\cdot10^{110},
 \label{Rinf}\end{equation}
where we used the estimations
\cite{Oikonomou:2025qub,OdintsovOS:2025} $M\approx3\cdot10^{22}$
eV,  $\Lambda\approx1.19\cdot10^{-66}$ eV${}^2$. We see in
Fig.~\ref{F1} that at ${\cal R}>{\cal R}_\mathrm{inf}$ the
parameter $F_R$ begins to grow as ${\cal R}/{\cal R}_\mathrm{inf}$
(whereas $F_R\simeq1$ if ${\cal R}<{\cal R}_\mathrm{inf}$),
$F_{RR}^*$ tends to the small constant $1/{\cal R}_\mathrm{inf}$
and $y$ evolves from small positive values at ${\cal R}<{\cal
R}_\mathrm{inf}$ to values
 $y\approx\frac{\cal R}{{\cal R}+{\cal R}_\mathrm{inf}}$ (close to 1, but $y<1$) at ${\cal R}>{\cal
 R}_\mathrm{inf}$.

The behavior of $x({\cal R})= 4{R F_{RRR}}/{F_{RR}}$ is shown also
in the bottom-right panel of Fig.~\ref{F1}. We see that at ${\cal
R}>10^{54}$ the value $x({\cal R})$ remains negative and tends to
zero if ${\cal R}\to\infty$. However at ${\cal R}<10^{52}$, this
parameter becomes strongly negative and satisfies the inequality
$x<-8$. This behavior satisfies the condition (\ref{yxineq}) $x<0$
and the more rigid condition  $-1<x<0$ \cite{Oikonomou:2025qub}
during the early-time acceleration.

We may conclude that for the logarithmic model (\ref{ModLog}), the
viability conditions (\ref{FRFRR}) and (\ref{yxineq}) are
fulfilled during all cosmological evolution for the best fit
solution and in its vicinity (if the restriction $y\le1$ is
applied only to the early-time inflationary epoch).

In two bottom-left panels we study the evolution of dark energy
density $\rho_{DE}(z)$ (\ref{rDE}) for the model (\ref{ModLog})
and the corresponding evolving EoS for dark energy. The dynamical
nature of $F(R)$ motivated dark energy can be measured via the
statefinder parameter $y_H (z)$
\cite{Hu:2007nk,Bamba:2012qi,reviews5},
\begin{equation} \label{yhdef}
    y_H(z)=\frac{\rho_{DE}(z)}{\rho_m^{0}}=\frac{H^2}{\Omega_m^0H_0^2}-(1+z)^3-X_r (1+z)^4 ,
\end{equation}
and two dark energy density parameters depicted in the
bottom-left panel:
\begin{equation}\label{OmegaDE}
    \Omega_{DE}(z)=\frac{\rho_{DE}(z)}{\rho_\mathrm{tot}(z)}=\frac{H^2-\frac{\kappa^2}3(\rho_m+\rho_r)}{H^2}
=\frac{y_H(z)}{y_H(z)+(z+1)^3 + X_r (z+1)^4}
\end{equation}
and
\begin{equation}\label{OmegaLDE}
    \Omega_{\Lambda}(z)=\frac{\rho_{DE}(z)}{\rho_\mathrm{tot}(0)}=\frac{H^2}{H_0^2}-\Omega_m^0(a^{-3}+X_ra^{-4})
=\Omega_m^0{y_H(z)}\,.
\end{equation}
The value $ \Omega_{DE}(z)$ measures the fraction of dark energy
during evolution at any redshift $z$, the parameter
$\Omega_{\Lambda}(z)$ equals constant ($\Omega_{\Lambda}$) for the
$\Lambda$CDM model, it shows how the considered $F(R)$ scenario
differs from $\Lambda$CDM. We see that $\Omega_{DE}(z)$
monotonously grows during its evolution, however for
$\Omega_{\Lambda}(z)$ the initial growth changes to a descent near
$z=1.6$.  Another measure of the difference between $F(R)$ and
$\Lambda$CDM scenarios is the dark energy EoS parameter expressed
as follows,
\begin{equation}\label{EoSDE}
    \omega_{DE}(z)=\frac{P_{DE}(z)}{\rho_{DE}(z)}=-1+\frac{z+1}{3y_H(z)}\frac{d
    y_H}{dz}\,.
\end{equation}
In the bottom panel of Fig.~\ref{F1} $\omega_{DE}(z)$ evolves the
$\Lambda$CDM value $\omega_{DE}=-1$ diminishing down to
$\approx-1.086$ near $z=3.75$ (the initial phantom stage), then
this parameter begins to grow and crosses the line $\omega=-1$
near $z=1.6$. This quintessence stage continues to $z=0$ with
growing up to $\omega_{DE}(0)\approx-0.85$. Such a behavior
supports the mentioned above analysis of Pantheon+ SNe Ia and BAO
DESI data with the dynamical dark energy models
\cite{Cai:2025mas,Ye:2024ywg,Chaudhary:2025vzy,Chaudhary:2025uzr,OdintsovSGS_game:2024,Giare:2024smz,Pan:2025qwy,Yang:2025mws},
in particular, for the $\omega_0\omega_a$CDM model with EoS
$\omega(z)=\omega_0+\omega_a\frac{z}{z+1}$ the obtained behavior
of $\omega_{DE}(z)$ corresponds to $\omega_0\approx-0.85$ and
negative $\omega_a$.

Calculating the $\chi^2$ function (\ref{chitot}) we analyze  the logarithmic model
(\ref{ModLog}) in confrontation its predictions with Pantheon SNe Ia, CC, CMB and BAO
DESI 2025 observational data. The results of our analysis  for pairs of free parameters
$\alpha$, $\gamma$, $H_0$, $\Omega_m^0$ and $\Omega_\Lambda$ are presented in
Fig.~\ref{F2} with contour plots at $1\sigma$ and $2\sigma$ confidence levels (CL) for
two-parameter distributions $\chi^2(\theta_j,\theta_k)$.

In these numerical calculations we use the approaches developed in
the previous papers
\cite{OdintsovSGS_game:2024,Odintsov:2025kyw,OdintsovSGS_Axi:2023,OdintsovSGS_LnAx:2024,OdintsovOS:2025,Odintsov:2025kyw}.
For for each pair of the chosen model parameters
$\theta_j,\theta_k$ we search the minimum of $\chi^2$ over the
other three parameters. In this procedure the grid spacing and
size of the box are determined at the initial stage, but the
center of the box is corrected and approximated during this
process. The prior ranges for the model parameters are limited
from their physical sense, in particular, for the model
(\ref{ModLog}) they are:
\begin{equation}\label{Priors1}
 \alpha\in[0,30];\quad\gamma\in[0,10];\quad
 \Omega_m\in[0.1,0.5];\quad \Omega_\Lambda\in[0.4,1];\quad H_0\in[50,100]\;\,\mathrm{km/s/Mpc}\,.
\end{equation}

In the bottom-left panel with contours in the $\Omega_m^0-H_0$
plane we compare the model (\ref{ModLog}) with the exponential
$F(R)$ model
\cite{OdintsovSGS:2017,CognolaENOSZ:2008,ElizaldeNOSZ:2011,Linder:2009}
 \begin{equation} \label{FRexpon}
 F(R)=R+F_\mathrm{inf}-2\Lambda\big(1-e^{-\beta{\cal R}}\big)\,.
  \end{equation}
In the top-right panel in Fig.~\ref{F2} we present one-parameter distributions
 $$ 
 \chi^2(H_0)=\min\limits_{\mathrm{other}\;\theta_j} \chi^2(\theta_1,\theta_2,\dots,H_0)\,.
 $$ 
for the mentioned two models and the $\Lambda$CDM model
(\ref{LCDM}) with,
\begin{equation} \label{HLCDM}
 H^2=H^2_0\big[\Omega_m^0(a^{-3}+ X_r a^{-4})+\Omega_\Lambda\big],\qquad\Omega_\Lambda=1-\Omega_m^0(1+ X_r)\,.
 \end{equation}
One can see that the logarithmic model (\ref{ModLog}) is the most
successful in its minimum of $\chi^2$ close to $2018.52$ that is
essentially lower than for the exponential (\ref{FRexpon}) and
$\Lambda$CDM (\ref{HLCDM}) models. These estimates of $\min\chi^2$
and the best fitted values of model parameters are shown in
Table~\ref{TMod}.

The likelihood functions ${\cal L}(\theta_j)$ for parameters $\theta_j$ in Fig.~\ref{F2}
are related with the one-parameter distributions $\chi^2(\theta_j)$:
   \begin{equation}
{\cal L}(\theta_j)= \exp\bigg[- \frac{\chi^2(\theta_j)-m^\mathrm{abs}}2\bigg]\ ,
 \label{likeli} \end{equation}
where $m^\mathrm{abs}$ the absolute minimum for $\chi^2$.
  \begin{figure}[th]
   \centerline{ \includegraphics[scale=0.68,trim=5mm 0mm 2mm -1mm]{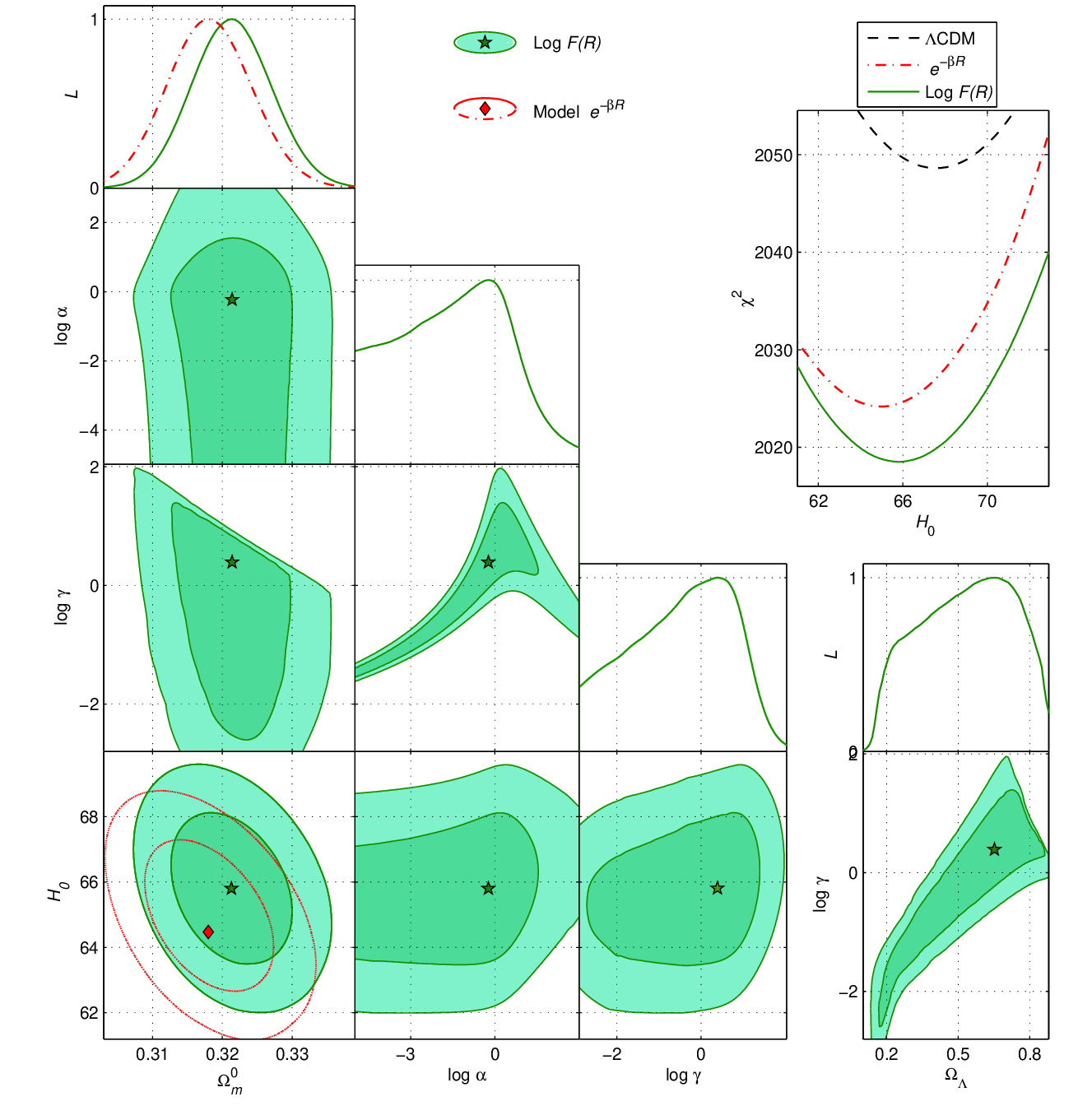}}
\caption{Contour plots of $\chi^2$ with $1\sigma$, $2\sigma$ CL,
likelihood functions $ {\cal L}(\theta_i)$ and one-parameter
distributions $\chi^2(H_0)$ for  the logarithmic model
(\ref{ModLog}) in comparison with the exponential (\ref{FRexpon})
and $\Lambda$CDM (\ref{HLCDM}) models for SNe Ia, CC, CMB and BAO
DESI data.}
  \label{F2}
\end{figure}

We see in  Fig.~\ref{F2} that for the model (\ref{ModLog}), small
values of $\alpha$ (and $\gamma$ to some extent) are included into
$1\sigma$ and $2\sigma$ CL domains. In the $\log\alpha-\log\gamma$
plane these suitable values form the long ``tail''. Such a
behavior is reflected in Table~\ref{TMod} where, for example, we
have the estimation $\alpha=0.80_{-0.794}^{+1.60}$.

For the logarithmic model (\ref{ModLog}), the best fit of the
Hubble parameter $H_0=65.81^{+1.51}_{-1.54}$ is larger from the
predicted value of the model (\ref{FRexpon}), but lower from the
predicted value of the $\Lambda$CDM scenario. Predictions of these
models for their common parameter are also different.

Fig.~\ref{F2} demonstrates the obvious large advantage the model
(\ref{ModLog}) in $\min\chi^2$ if we compare it with the other two
scenarios, namely the $\Lambda$CDM and the model of Eq.
(\ref{FRexpon}). This advantage does not vanish even when we
consider the number of free parameters $N_p$ for each model
following the Akaike information criterion (AIC) and the Bayesian
information criterion (BIC) \cite{Liddle_ABIC:2007},
 \begin{equation}
 \mbox{AIC} = \min\chi^2 +2N_p\,,\qquad \mathrm{BIC} = \min\chi^2 +N_p\cdot\log(N_d)\;..
  \label{AICBIC}\end{equation}
Here $N_d=1744$ is the number of data points, $N_p=5$, 4 and 2 for
the models  (\ref{ModLog}), (\ref{FRexpon}) and $\Lambda$CDM
respectively. The AIC and BIC estimates for the considered models
are presented in Table~\ref{TMod}. We see that information
criteria (\ref{AICBIC}) support the advantage of the logarithmic
model (\ref{ModLog}). However the model (\ref{ModExp}) with the
exponent $e^{\varepsilon{\cal R}^{-\beta}}$ also demonstrates
attractive results in Table~\ref{TMod}. This model is considered
in detail in the next section.
\begin{table}[ht]\label{TMod}
\begin{center}
\caption{Best fits with $1\sigma$ errors, $\min\chi^2$, AIC, BIC  from SNe Ia, $H(z)$,
CMB and BAO DESI DR2 data for the logarithmic model  (\ref{ModLog}), the model
(\ref{ModExp}) with $e^{\varepsilon{\cal R}^{-\beta}}$ in comparison with the model
(\ref{FRexpon})  with $e^{-\beta{\cal R}}$ and $\Lambda$CDM model (\ref{HLCDM}).}
\begin{tabular}{|l|c|c|c|c|c|c|c|}  \hline
 \hline  Model &   $\min\chi^2/d.o.f$& AIC & BIC& $\Omega_m^0$& $H_0$& $\Omega_\Lambda$&  other parameters \\
\hline
 Log (\ref{ModLog}) & 2018.52 /1739 & 2028.52& $2055.84$ &$0.3213^{+0.0057}_{-0.0058}$
& $65.81^{+1.51}_{-1.54}$  & $0.625^{+0.41}_{-0.38}$ &$\alpha=0.80_{-0.794}^{+1.60}$, $\gamma=1.483_{-1.293}^{+1.565}$ \rule{0pt}{1.1em}  \\
\hline
 (\ref{ModExp}):\,$e^{\varepsilon{\cal R}^{-\beta}}$ & 2018.50 /1739 & 2028.50&$2055.82$&$0.3212^{+0.0068}_{-0.0058}$
& $65.66^{+1.54}_{-1.52}$ & $0.629^{+0.117}_{-0.404}$ &$\beta=1_{-0}^{+0.36}$, $\varepsilon=1.02^{+2.52}_{-0.75}$ \rule{0pt}{1.1em}  \\
\hline
 (\ref{FRexpon}):\,$e^{-\beta{\cal R}}$ & 2024.17 /1740 & 2032.17& $2054.03$
&$0.3180^{+0.0061}_{-0.0060}$
& $64.46^{+1.53}_{-1.52}$  & $0.5645^{+0.010}_{-0.006}$ &$\beta=0.707^{+0.102}_{-0.075}$  \rule{0pt}{1.1em}  \\
\hline
$\Lambda$CDM& 2048.62 /1742 & 2052.62& 2063.55 & $0.2923^{+0.0011}_{-0.0012}$& $67.56^{+1.55}_{-1.52}$& - & -  \rule{0pt}{1.1em}  \\
\hline
  \hline \end{tabular}
 \end{center}
\end{table}

\section{Exponential $F(R)$ Model with $e^{\varepsilon{\cal R}^{-\beta}}$}
\label{Exp}

As mentioned in the previous sections, we consider here the $F(R)$
model (\ref{ModExp}) with the exponential factor
$e^{\varepsilon{\cal R}^{-\beta}}$ which satisfies the viability
conditions $F_{RR}>0$ and $x<0$ during all the cosmological
evolution eras. However, the conditions $F_{R}>0$ and $y>0$ for
this model need verification, because they may be violated at late
times if the term $\varepsilon\beta{\cal R}^{-\beta}$ appear to be
larger than 1 at some values ${\cal R}$ (see Table~\ref{T1}). For
the model (\ref{ModExp}) as for the previous $F(R)$ scenario
(\ref{ModLog}), we should integrate the system of equations
(\ref{eqH1}), (\ref{eqR1}) and investigate its solutions for
solving the viability problem.

Since the model (\ref{ModExp}) has the  $\Lambda$CDM-like
asymptotic behavior with $F(R)\to R -2\Lambda+F_\mathrm{inf}$ if
$R\to\infty$, we also use the $\Lambda$CDM-like asymptotic
conditions (\ref{asymLCDM}) at the initial point $a_\mathrm{ini}$.
The value $a_\mathrm{ini}$ we also determine from the condition
$2\Lambda F_{RR}=\delta$, where $\delta$ is a small value of order
$10^{-10}$. This condition for the model (\ref{ModExp}) may be
reduced to the equations,
 \begin{equation} \label{ainiE}
{\cal R}_\mathrm{ini} = \left\{\begin{array}{ll}(\varepsilon^2/\delta)^{1/3},&  \beta=1,\\
\big[\varepsilon\beta(\beta-1)/\delta\big]^{1/(1+\beta)}, &\beta>1.\end{array} \right.
 \qquad a_\mathrm{ini}= \bigg[\frac{2\Omega_\Lambda^*}{\Omega_m^*}({\cal R}_\mathrm{ini}-2)\bigg]^{-1/3}\;,
 \end{equation}
Starting from $a_\mathrm{ini}$ we integrate the system, including
Eq.~(\ref{eqH2}), and the equations (\ref{eqR1}) for this model
(\ref{ModLog}) in the form,
\begin{equation}
\frac{d{\cal R}}{d\log a}={\cal R} \frac{\big[\Omega_m^{*}(a^{-3}+ X_r a^{-4})
-\Omega_\Lambda^{*}(\frac\nu2+\varepsilon\beta{\cal R}^{1-\beta}e^{\varepsilon{\cal
R}^{-\beta}})\big]/{E^2} -e^{\varepsilon{\cal R}^{-\beta}}(1-\varepsilon\beta{\cal
R}^{-\beta})}
 {\varepsilon\beta\big[(\beta-1){\cal R}^{-\beta}+\varepsilon\beta{\cal R}^{-2\beta}\big]
 \,e^{\varepsilon{\cal R}^{-\beta}}}
  \label{eqExpR2}
  \end{equation}
instead of Eq.~(\ref{eqR2}). The results of calculations are
confronted with the same set of observational data including
Pantheon+SNe Ia, BAO DESI 2025, $H(z)$ (CC) and CMB data,
described in Appendix. In Fig.~\ref{F3} we analyze the viability
conditions (\ref{FRFRR}), (\ref{yxineq}).

In the top-left panel of  Fig.~\ref{F3} the $1\sigma$ and
$2\sigma$ contour plots of $\chi^2$ are depicted in
$\beta-\log\varepsilon$ plane with the contour  $F_{R}=0$ (the
blue line). Here we observe a very unusual  behavior of the model
(\ref{ModExp})  with $e^{\varepsilon{\cal R}^{-\beta}}$: the
corresponding  $\chi^2$ function (\ref{chitot}) achieves the
absolute minimum $m^\mathrm{abs}=\min\chi^2\approx2018.50$ if
$\beta=1$ ($\beta=1_{-0}^{+0.36}$, the square in the panel),
however this $\chi^2$ has the local minimum $\approx 2018.68$
(denoted as the hexagram) at more high $\beta\approx 2$, more
precisely, $\beta=2.035_{-0.345}^{+0.256}$. At some values $\beta$
and $\varepsilon$ between the mentioned minima points we see the
white domain with large values of $\chi^2$. Another ``white''
domain lies at small $\beta$ and large $\varepsilon$. In these
domains the model (\ref{ModExp}) appears to be unsuccessful.

In addition, in the last domain with small $\beta$ and large
$\varepsilon$, the viability condition  (\ref{FRFRR}) $F_R>0$ (for
all $R$) is violated, moreover, the non-physical domain with
$F_R<0$ includes also some area with $1<\beta<1.3$ and
$\varepsilon>2.4$, where the $\chi^2$ values are acceptable. Note
that in the case $\beta=1$ the condition $F_R>0$ is fulfilled.

However, if we exclude the domain with $\min F_R<0$ surrounded by
the blue $\min F_R=0$ line in Fig.~\ref{F3}, in the remaining
domain the model (\ref{ModExp}) works successfully and the
viability conditions $F_R>0$, $F_{RR}>0$, $y>0$, $x<0$ are
fulfilled that can be seen in other panels of Fig.~\ref{F3}.
  \begin{figure}[th]
   \centerline{ \includegraphics[scale=0.68,trim=5mm 0mm 2mm -1mm]{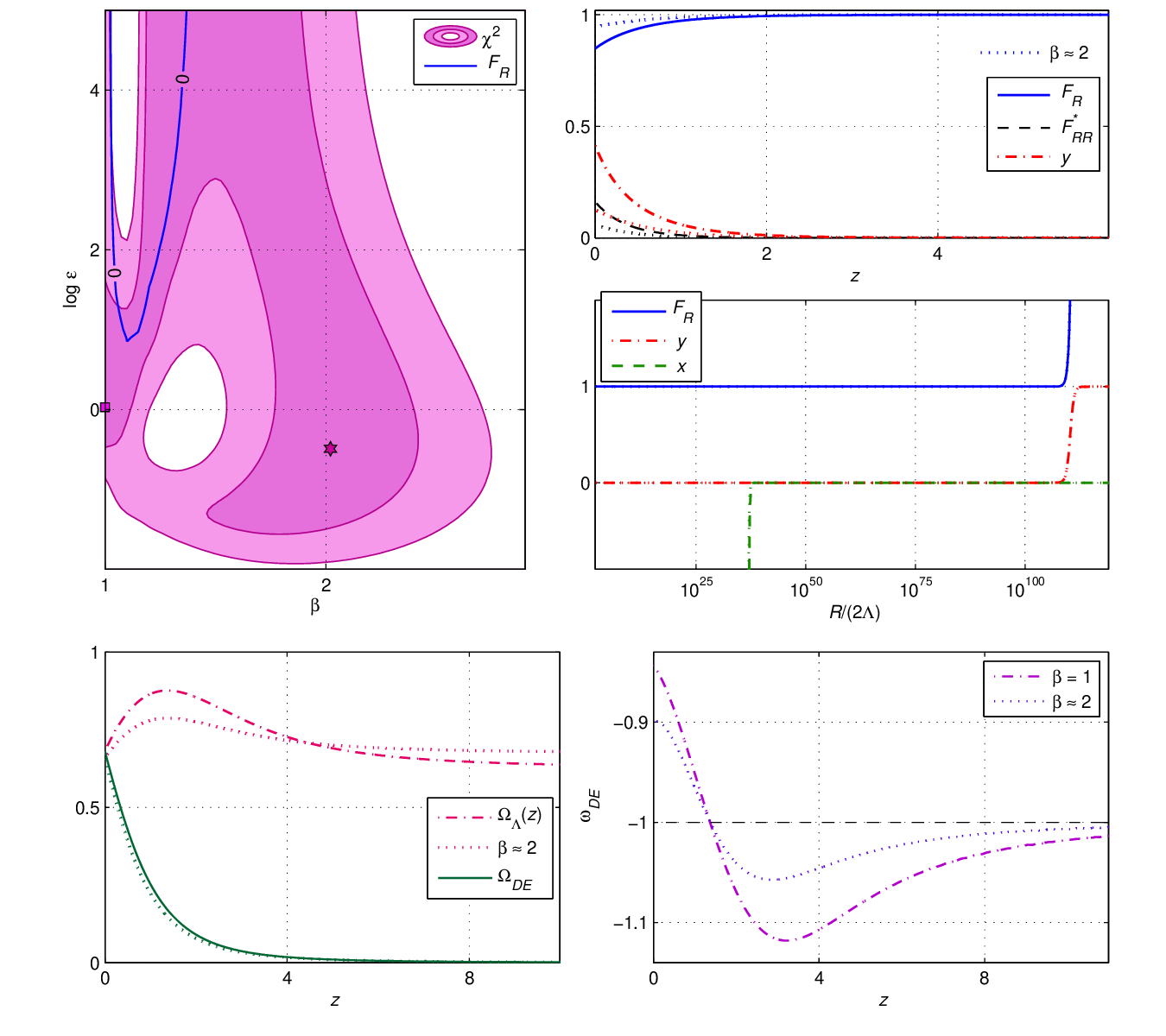}}
\caption{Contour plots of $\chi^2$ with $1\sigma$, $2\sigma$ CL and  $F_R=0$ in the
$\beta-\log\varepsilon$ plane for the model (\ref{ModExp})  with $e^{\varepsilon{\cal
R}^{-\beta}}$ (the top-left panel); evolution of $F_R$, $F_{RR}^*=2\Lambda F_{RR}$, $y$
and $x$ az functions of redshift $z$ (late-time) and the Ricci scalar ${\cal
R}=R/(2\Lambda)$ (early-time dynamics) in the top-right panels; the dark energy density
and EoS parameters in the bottom panels. }
  \label{F3}
\end{figure}
In particular, in the top-right panels of Fig.~\ref{F3} the
evolution of $F_R$, $F_{RR}^*=2\Lambda F_{RR}$, $y$ and $x$ is
depicted as functions of the redshift $z$ (at late time) and as
functions of the Ricci scalar ${\cal R}=R/(2\Lambda)$ at
early-time dynamics. The model parameters  $\beta=1$,
$\varepsilon=1.02$ and $\Omega_i$ from Table~\ref{TMod} correspond
to the global minimum of $\chi^2$ for the solid, dashed and
dash-dotted lines. The dotted lines describe the behavior of the
same functions for the local minimum with  $\beta=2.035$,
$\varepsilon==0.613$ denoted as the hexagram.

We see, that the functions  $F_R$, $F_{RR}$, $y$ and $x$ satisfy
the viability conditions (\ref{FRFRR}), (\ref{yxineq}) and behave
similarly to the same functions for the logarithmic model
(\ref{ModLog}) in Fig.~\ref{F1}. For both models $F_R(z)$ achieves
its minimum at $z=0$ (it is positive in the domain of viability),
the function $F_R(R)$ begins to grow, if the normalized Ricci
scalar  ${\cal R}$  is larger ${\cal
R}_\mathrm{inf}={M^2}/(4\Lambda)\approx1.9\cdot10^{110}$. If
${\cal R}$ grows over ${\cal R}_\mathrm{inf}$ the parameter $y(R)$
transfers from small positive to close to 1 values. The expression
$F_{RR}^*$ at high ${\cal R}$ is positive and very small. The
parameter $x({\cal R})= 4{R F_{RRR}}/{F_{RR}}$ for the model
(\ref{ModExp}) at ${\cal R}>10^{37}$ is negative and tends to zero
if ${\cal R}\to\infty$, but at ${\cal R}<10^{35}$ the value r
$x({\cal R})$ is close to $-12$, it remains negative during all
the evolution. The early time behavior of these parameters for two
considered cases with $\beta=1$ and $\beta=2.035$ is very similar.

In the two bottom panels of Fig.~\ref{F3} the evolution of dark
energy density parameters
$\Omega_{\Lambda}(z)={\rho_{DE}(z)}/{\rho_\mathrm{tot}(0)}$
(\ref{OmegaLDE}),
$\Omega_{DE}(z)={\rho_{DE}(z)}/{\rho_\mathrm{tot}(z)}$
(\ref{OmegaDE}) and the  dark energy EoS parameter $
\omega_{DE}(z)=P_{DE}/\rho_{DE}$ (\ref{EoSDE})  for the model
(\ref{ModExp}) is shown for the mentioned cases of $\chi^2$
minima: $\beta=1$ and $\beta\approx2$. The dotted curves also
correspond to the case $\beta\approx2$. The dark energy density
parameters reflect variations of the dark energy density
$\rho_{DE}(z)$. We may conclude that the  $F(R)$ model
(\ref{ModExp})  with $e^{\varepsilon{\cal R}^{-\beta}}$ similarly
to the previous scenario (\ref{ModLog}) behaves at late times as a
dynamical dark energy $F(R)$ model.

The dynamical behavior for this case, may be seen from the
evolution of the dark energy EoS $\omega_{DE}(z)$ in the
bottom-right panel of Fig.~\ref{F3}. From the initial $\Lambda$CDM
value $\omega_{DE}=-1$ this parameter diminishes, and this phantom
stage ends at $z\approx 1.37$, when the parameter $\omega_{DE}$
crosses the line $\omega=-1$. Further, the quintessence stage
continues to $z=0$ with growing up to
$\omega_{DE}(0)\approx-0.843$. Thus, the $F(R)$ model
(\ref{ModExp}) like the logarithmic scenario (\ref{ModLog})
describes the Pantheon+ SNe Ia and BAO DESI observational data as
the dynamical dark energy models with varying EoS, for example,
the $\omega_0\omega_a$CDM model
\cite{Chaudhary:2025vzy,Chaudhary:2025uzr}.
  \begin{figure}[th]
   \centerline{ \includegraphics[scale=0.72,trim=5mm 0mm 2mm -1mm]{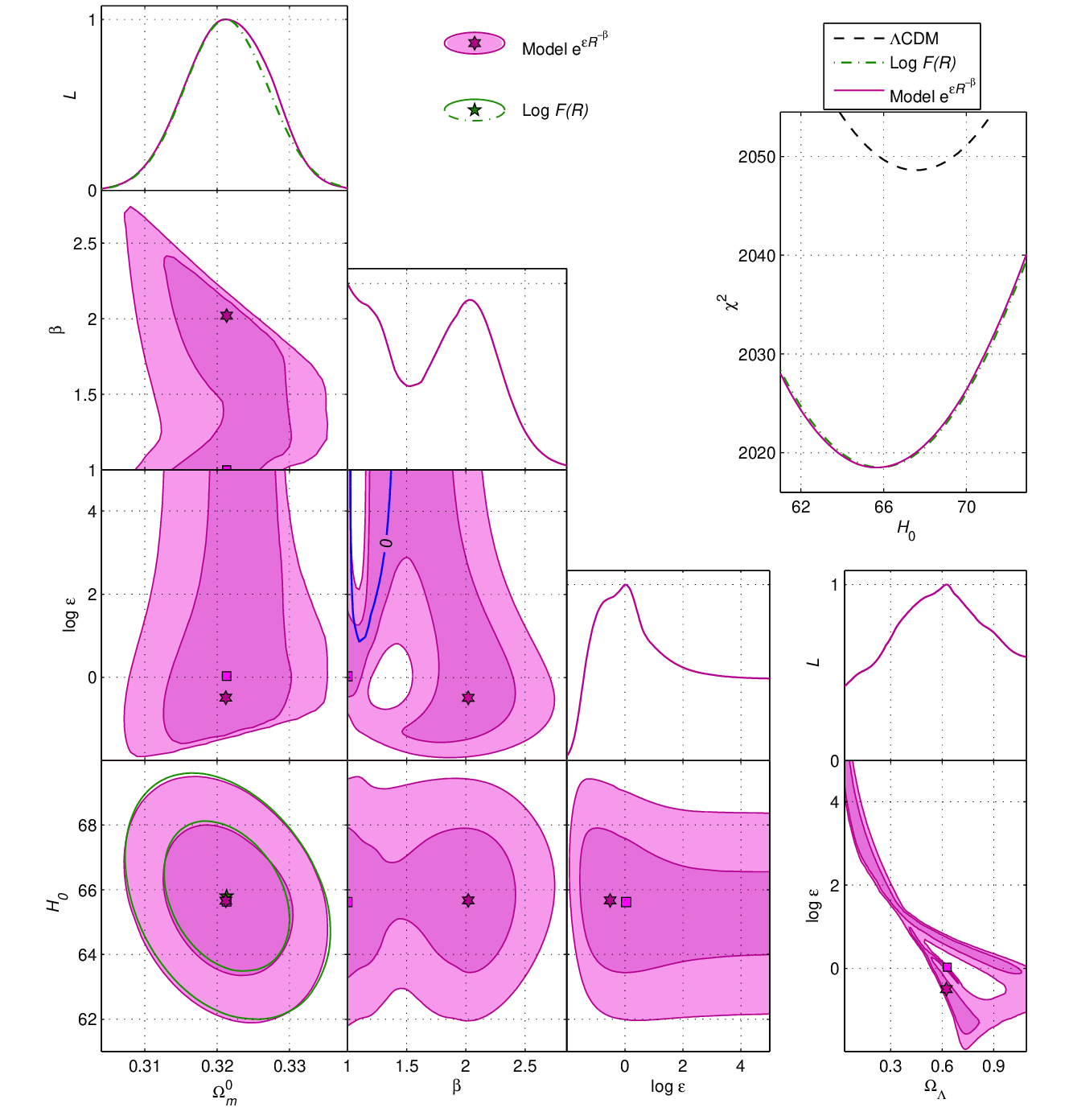}}
\caption{For the model (\ref{ModExp})  with $e^{\varepsilon{\cal R}^{-\beta}}$ the
contour plots at $1\sigma$, $2\sigma$ CL, likelihood functions $ {\cal L}(\theta_i)$ and
one-parameter distributions $\chi^2(H_0)$ are shown in comparison with the logarithmic
(\ref{ModLog}) and $\Lambda$CDM (\ref{HLCDM}) models for SNe Ia, CC, CMB and BAO DESI
data. }
  \label{F4}
\end{figure}
In Fig.~\ref{F4} we reproduce the detailed analysis of the $F(R)$
model (\ref{ModExp}) with $e^{\varepsilon{\cal R}^{-\beta}}$
including likelihoods and contour plots for two-parameter
distributions in planes with pairs of model parameters. Here the
mentioned above minima points of $\chi^2$ are shown like in
Fig.~\ref{F4}, where the squares denote the absolute minimum.

In the bottom-left panel with contours in the $\Omega_m^0-H_0$ plane and also in the
panels with ${\cal L}(\Omega_m^0)$, and $ \chi^2(H_0)$ we compare this model with the
previous scenario (\ref{ModLog}) and with the $\Lambda$CDM scenario in the top-right
panel. We see here and in Table~\ref{TMod} that the best fits for $\Omega_m^0$ and
$H_0$, and also the contour plots for $\chi^2(\Omega_m^0,H_0)$ for both considered
$F(R)$ models (\ref{ModExp}) and (\ref{ModLog}) are very close. The absolute minima
$m^\mathrm{abs}=\min\chi^2$ are also close, they are 2018.52 and 2018.50 respectively,
that is essentially better than for the $\Lambda$CDM model. This large advantage does
not vanish, if we consider the Akaike and Bayesian information criteria (\ref{AICBIC})
\cite{Liddle_ABIC:2007}.

The above mentioned ``white'' domains with large $\chi^2$ values
can be observed not only in $\beta-\log\varepsilon$, but also in
$\Omega_\Lambda-\log\varepsilon$ plane of Fig.~\ref{F4}. Remind
that the square and the hexagram denote the best fit points of
$\chi^2$ achieved at $\beta=1$ and near $\beta=2$ respectively.
The best fits of $\Omega_\Lambda$ for these points appeared to be
close, but $\varepsilon$ are different.

The described above two local minima of $ \chi^2$ at $\beta=1$ and
$\beta\approx 2.035$ may be seen in the  likelihood ${\cal
L}(\beta)$, the intermediate values of $\beta$ are less
successful. For other parameters we do not see such a separation,
their best fits for both minima appear to be neighboring or
coinciding (for $\Omega_m^0$ and $H_0$).

Large advantage in AIC for both considered $F(R)$ scenarios (\ref{ModExp}) and
(\ref{ModLog}) in comparison to  $\Lambda$CDM model supports our previous results for
other $F(R)$ models with  $\Lambda$CDM-like asymptotic behavior at high $R$
\cite{OdintsovSGS_game:2024,OdintsovOS:2025}. This success of $F(R)$ models is connected
with their capability to mimic the dynamical dark energy behavior that is necessary for
describing the  Pantheon+ SNe Ia and BAO DESI observational data
\cite{PantheonP:2022,DESI:2025zgx}.

\section{Conclusions}

In this paper we explored two $F(R)$ gravity models: the model (\ref{ModLog}) with the
logarithmic term $\gamma\log(\alpha{\cal R})$ (where ${\cal R}=\frac{R}{2\Lambda}$) and
the model (\ref{ModExp}) with the exponential term $e^{\varepsilon{\cal R}^{-\beta}}$ in
its Lagrangian. These scenarios have the $\Lambda$CDM-like asymptotic behavior in the
large $R$ limit, they are extracted from more wide classes of  $F(R)$ gravities
considered previously in the papers \cite{Oikonomou:2025qub,OdintsovOS:2025}. Both
scenarios (\ref{ModExp}) and (\ref{ModLog}) provide a unified description of early-time
inflationary epoch and late-time acceleration driven by some form of dynamical dark
energy, generated in $F(R)$ dynamics.

Both models (\ref{ModExp}) and (\ref{ModLog}) appeared to be very successful in
describing the observational data from Pantheon+ SNe Ia \cite{PantheonP:2022}, BAO DESI
DR2 \cite{DESI:2025zgx}, the CC Hubble parameter measurements $H(z)$ and CMB data
\cite{Planck2018}: they are advantageous over the $\Lambda$CDM scenario (\ref{HLCDM})
and the exponential model (\ref{FRexpon}) with $F_{DE}(R)=-2\Lambda\big(1-e^{-\beta{\cal
R}}\big)$ not only in terms of $\min\chi^2$ but also in the information criteria AIC and
BIC, as shown in Table~\ref{TMod}. Moreover, the models (\ref{ModExp}) and
(\ref{ModLog}) are more successful in their $\min\chi^2$, AIC and BIC results than other
$F(R)$ models, explored in Ref.~\cite{OdintsovOS:2025}.

However, unlike the $\Lambda$CDM and exponential (\ref{FRexpon}) scenarios, the models
(\ref{ModExp}) and (\ref{ModLog}) have some problems with the viability conditions
(\ref{FRFRR}) and (\ref{yxineq}) for the parameters $F_{R}$, $F_{RR}$,
$y=RF_{RR}/F_{R}$, $x= 4{R F_{RRR}}/{F_{RR}}$. The most dangerous problems take place
with the condition $F_{R}>0$ (excluding antigravity effects) that can be violated during
the late-time evolution. To investigate these conditions, we analyzed not only the
Lagrangian model parameters $\gamma$, $\alpha$, $\varepsilon$, $\beta$, but also the
parameters $\Omega_m^0$, $\Omega_\Lambda$ (\ref{Omega_mL}), which determine an evolution
of the Ricci scalar $R(z)$. As the result of this analysis we observe for both models in
Figs.~\ref{F1} and \ref{F3} the forbidden domains in the parameter spaces, where $\min
F_{R}<0$. For the logarithmic model (\ref{ModLog}) this forbidden domain (with small
$\alpha$ and large $\gamma$) lies far from the best fit values of model parameters and
their $2\sigma$ vicinity. But for the model (\ref{ModExp}) the forbidden domain with
$F_{R}<0$ occupies some area with suitable values of $\chi^2$ in $1\sigma$ and $2\sigma$
CL domains, that can be seen in Fig.~\ref{F3}. This area should be excluded from
cosmological applications. Fortunately, in the remaining domain with $\min F_R>0$ the
model (\ref{ModExp}) works successfully. For both models (\ref{ModExp}) and
(\ref{ModLog}) other viability conditions (\ref{FRFRR}), (\ref{yxineq}) $F_{RR}>0$,
$y>0$, $x<0$ are fulfilled during all early-time and late-time evolution (see
Figs.~\ref{F1}, \ref{F3}).

The model (\ref{ModExp}) with  $e^{\varepsilon{\cal R}^{-\beta}}$ achieves its best fit
if the parameter $\beta$ equals 1. So we can consider its narrowed variant with
$\beta=1$:
\begin{equation}\label{ModExp1}
F(R)=\frac{R^2}{M^2}+ R\,e^{\varepsilon/{\cal R}}-2(1+\varepsilon)\,\Lambda\,.
\end{equation}
This model has $N_p=4$ free model parameters, hence its best
results in the information criteria (\ref{AICBIC}) become better:
AIC${}\approx2026.50$ and BIC${}\approx2048.36$. Thus, the
narrowed model (\ref{ModExp1}) has the additional advantage over
the $\Lambda$CDM model.

From Table~\ref{TMod} we may conclude that the large advantage of
the $F(R)$ models (\ref{ModExp}) and (\ref{ModLog}) over the
$\Lambda$CDM scenario in terms of $\min\chi^2$, AIC and BIC is
connected, in particular, with the fact, that the best fits of
these  $F(R)$ models for $\Omega_m^0$ and $H_0$ (very close to
each other) are  far from their $\Lambda$CDM best fits. For the
Hubble constant  the $\Lambda$CDM best fit
$H_0=67.56^{+1.55}_{-1.52}$ is more than $1\sigma$ larger, but for
$\Omega_m^0$ it is more than $3\sigma$ less in comparison to both
$F(R)$ models.

This difference in the best fits and also in achieved $\min\chi^2$
takes place also for other $F(R)$ scenarios, confronted with
Pantheon+ SNe Ia and  DESI BAO observational data: for the
exponential model (\ref{FRexpon}), for its generalization  with
$$F(R)=R+F_\mathrm{inf}(R)-\Lambda\big(2-\alpha e^{-\varepsilon{\cal R}}\big)$$
 and the model with
$$F(R)=R+F_\mathrm{inf}(R)-\frac{2\Lambda}{1+\alpha e^{-\varepsilon{\cal R}}}\,,$$
 considered in Ref.~\cite{OdintsovOS:2025}.

We may conclude that $F(R)$ gravities, in particular, the models
(\ref{ModExp}), (\ref{ModLog}) have the mentioned large advantage
over the $\Lambda$CDM scenario, because they are capable to mimic
the dynamical dark energy with suitable behavior of its density
$\rho_{DE}(z)$ (\ref{rDE}) and the dark energy EoS parameter
$\omega_{DE}(z)=P_{DE}/\rho_{DE}$ (\ref{EoSDE}). The evolution of
these parameters for the best fit solutions in scenarios
(\ref{ModLog}) and (\ref{ModExp}) is shown in Figs.~\ref{F1},
\ref{F3}. We see the initial phantom stage that at $z\simeq 1.5$
transfers into the quintessence stage continuing up to the present
time. This behavior my be also described in the framework of
numerous dynamical dark energy models with EoS $\omega=\omega(z)$
\cite{Cai:2025mas,Ye:2024ywg,Chaudhary:2025vzy,Chaudhary:2025uzr,OdintsovSGS_game:2024,Giare:2024smz,Pan:2025qwy,Yang:2025mws},
in particular, with $\omega_0\omega_a$CDM model where
$\omega(z)=\omega_0+\omega_a\frac{z}{z+1}$,  the obtained  $F(R)$
behavior of $\omega_{DE}(z)$ corresponds to
$\omega_0\in[-0.9,-0.84$ and negative $\omega_a$.

An important aspect of the analysis performed in this work is the
physical interpretation of the observationally allowed regions of
the models parameter space. The viability conditions of $f(R)$
gravity already impose some non-trivial theoretical constraints,
with $F_R>0$ and $F_{RR}>0$, and also the existence of a stable
high-curvature regime that correctly reproduces GR at early times.
These conditions considerably restrict the allowed parameter space
before any observational constraints are actually applied.

The regions favored by the cosmological observations are well
within the theoretically viable domain, and are not located near
pathological boundaries of the model. In particular, the
observationally preferred parameter intervals correspond to models
that closely approach the $\Lambda$CDM model at high redshifts,
while allowing small, but still detectable late-time deviations
driven by the $f(R)$ gravity. This demonstrates that the
successful fits are not the result of some fine tuning near the
excluded regions, but it is rather imposed in a physically
well-behaved and theoretically allowed sector of the total
parameter space.

\section*{Appendix}
\label{Observ}

In this paper, we follow the previous works
\cite{OdintsovSGS_game:2024,OdintsovOS:2025,Odintsov:2025kyw} and include in our tests
the following observational data: (a) Type Ia Supernovae (SNe Ia) data from the
Pantheon+ sample database, (b) estimations of the Hubble parameter $H(z)$ or Cosmic
Chronometers (CC), (c) parameters from the Cosmic Microwave Background radiation (CMB)
and the  Baryon Acoustic Oscillations (BAO) data from Dark Energy Spectroscopic
Instrument (DESI) collaboration 2025 \cite{DESI:2025zgx}. For SNe Ia data we use the
Pantheon+ catalogue \cite{PantheonP:2022} with $N_{\mathrm{SN}}=1701$ datapoints of the
distance moduli $\mu_i^\mathrm{obs}$ at redshifts $z_i$ and calculate the $\chi^2$
function:
$$
\chi^2_{\mathrm{SN}}(\theta_1,\dots)=\min\limits_{H_0} \sum_{i,j=1}^{N_\mathrm{SN}}
 \Delta\mu_i\big(C_{\mathrm{SN}}^{-1}\big)_{ij} \Delta\mu_j\ ,\qquad 
 \Delta\mu_i=\mu^\mathrm{th}(z_i,\theta_1,\dots)-\mu^\mathrm{obs}_i\ .
 $$
with the covariance matrix $C_{\mbox{\scriptsize SN}}$ \cite{PantheonP:2022} and
 theoretical estimates:
\begin{equation}
 \mu^\mathrm{th}(z) = 5 \log_{10} \frac{(1+z)\,D_M(z)}{10\mbox{pc}},\qquad D_M(z)= c \int\limits_0^z\frac{d\tilde z}{H(\tilde
 z)}.    \label{muDM}
\end{equation}

For the Hubble parameter data $H(z)$ we work here with $N_H=32$ datapoints of
$H^\mathrm{obs}(z_i)$ (Cosmic Chronometers) used earlier in the previous papers
\cite{OdintsovSGS_Axi:2023,OdintsovSGS_LnAx:2024,OdintsovSGS_game:2024,OdintsovOS:2025}.
 The corresponding  $\chi^2$ function  yields:
 $$
\chi^2_{H}= \sum_{i=1}^{N_H} \left[\frac{H^\mathrm{obs}(z_i)
 -H^\mathrm{th}(z_i; \theta_k)}{\sigma_{H,i}}\right]^2 \, .
 $$
The CMB observational parameters in accordance with
Refs.~\cite{OdintsovSGS_game:2024,OdintsovOS:2025} are used here as the set
\cite{Planck2018}
 $$
\mathbf{x}=\left(R,\ell_A,\omega_b \right)\, ,\quad
R=\sqrt{\Omega_m^0}\frac{H_0D_M(z_*)}c\, ,\quad \ell_A=\frac{\pi D_M(z_*)}{r_s(z_*)}\, ,
\quad\omega_b=\Omega_b^0h^2
  $$
with the data priors \cite{ChenHW:2018}
 $$
\mathbf{x}^\mathrm{Pl}=\left( R^\mathrm{Pl},\ell_A^\mathrm{Pl},\omega_b^\mathrm{Pl}
\right) =\left( 1.7428\pm0.0053,\;301.406\pm0.090,\;0.02259\pm0.00017 \right)
$$ for scenarios with zero spatial curvature
and  $\Lambda$CDM-like asymptotic behavior. The comoving sound horizon  $r_s(z_*)$ is
calculated as the integral
\cite{OdintsovSGS_Axi:2023,OdintsovSGS_LnAx:2024,OdintsovSGS_game:2024}:
  \begin{equation}
r_s(z)=  \int_z^{\infty} \frac{c_s(\tilde z)}{H (\tilde z)}\,d\tilde
z=\frac1{\sqrt{3}}\int_0^{1/(1+z)}\frac{da}
 {a^2H(a)\sqrt{1+\big[3\Omega_b^0/(4\Omega_\gamma^0)\big]a}}\ ,
  \label{rs2}\end{equation}
where the redshift $z_*$ related to the photon-decoupling epoch is estimated following
Refs.~\cite{OdintsovSGS_Axi:2023,ChenHW:2018}. We calculate the $\chi^2$ function with
the covariance matrix $C_{\mathrm{CMB}}=\| \tilde C_{ij}\sigma_i\sigma_j \|$
\cite{ChenHW:2018}:
 $$
\chi^2_{\mathrm{CMB}}=\min_{\omega_b,H_0}\Delta\mathbf{x}\cdot
C_{\mathrm{CMB}}^{-1}\left( \Delta\mathbf{x} \right)^{T}\, ,\quad \Delta
\mathbf{x}=\mathbf{x}-\mathbf{x}^\mathrm{Pl}\,.
 $$
For the BAO we use the new DESI data from Data Release 2
\cite{DESI:2025zgx}.  We calculate and compare with measurements
the values,
 $$
\frac{D_M(z)}{r_d},\qquad\frac{D_H(z)}{r_d}=\frac{c}{H(z)\,r_d},\qquad
\frac{D_V(z)}{r_d}=\frac{(zD_H D_M^2)^{1/3}}{r_d},
 $$
where $r_d=r_s(z_d)$  is calculated as the integral (\ref{rs2})
and $z_d$ being the redshift at the end of the baryon drag era,
estimated by the Planck 2018 data \cite{Planck2018}. We use the
observational value $D_V(z_1)/r_d$ at $z_1=0.295$ and data points
with $D_M(z_i)/r_d$ and $D_H(z_i)/r_d$ for higher redshifts $z_i$
available in Ref.~\cite{DESI:2025zgx}. The corresponding $\chi^2$
function is,
 $$
\chi^2_{\mathrm{BAO}}(\theta_1,\dots)=\bigg[\frac{\Delta_V(z_1)}{\sigma_V(z_1)}\bigg]^2
+\sum_{i=2}^8 [\Delta_M(z_i)\;\,\Delta_H(z_i)]\,C^i_{M,H}
\bigg[\begin{array}{c}\!\Delta_M(z_i)\!\\ \Delta_H(z_i)\end{array}\bigg],
 $$
where,
$\Delta_q=\big(\frac{D_q}{r_d}\big)^\mathrm{th}-\big(\frac{D_q}{r_d}\big)^\mathrm{obs}$
with $q=V,\,M,\,H$; $C^i_{M,H}$ are the covariance matrices,
including the errors $\sigma_q(z_1)$ and the cross-correlation
coefficients  $r^i_{M,H}$ between $D_M(z_i)/r_d$ and
$D_H(z_i)/r_d$.

\end{document}